\documentclass[12pt]{article}
\usepackage{amsmath,amssymb,amsthm,amsxtra,overpic,bbm,bm,epsfig,ulem}
\usepackage{color}
\usepackage{cite}

\usepackage{hyperref}
\usepackage{url}

\def\thefootnote{\fnsymbol{footnote}}

\begin{document}

\vspace{0.2cm}

\begin{center}
{\large\bf Quantum-entangled $B$ mesons and CP violation: a brief
overview}
\end{center}

\vspace{0.2cm}

\begin{center}
{\bf Zhi-zhong Xing$^{1,2,3}$}
\footnote{E-mail: xingzz@ihep.ac.cn}
\\
{\small $^{1}$Institute of High Energy Physics, Chinese Academy of Sciences,
Beijing 100049, China} \\
{\small $^{2}$School of Physical Sciences,
University of Chinese Academy of Sciences, Beijing 100049, China} \\
{$^{3}$Center of High Energy Physics, Peking University, Beijing 100871, China}
\end{center}

\vspace{1cm}

\begin{abstract}
This paper is intended to provide a brief overview of the quantum-entangled
production of neutral meson-antimeson pairs, and to highlight the first
discovery of a clean physical CP-violating phase in coherent neutral-$B$
decays as a smoking gun of the Kobayashi-Maskawa mechanism of CP violation
in the standard model of particle physics.
\end{abstract}

%\keywords{quantum entanglement; meson-antimeson mixing; $B$ decays, CP violation.}
%\ccode{PACS numbers: 03.65.Ud, 11.30.Er, 12.15.-y, 13.25.Hw}

\newpage

\def\thefootnote{\arabic{footnote}}
\setcounter{footnote}{0}

\section{Quantum entanglement: from a hypothesis to reality}	

Just as Henri Poincar$\rm\acute{e}$ once said, ``Hypotheses are what we
lack the least", especially for a theorist. The famous Einstein-Podolsky-Rosen (EPR)
quantum entanglement was simply a theoretical hypothesis when it was proposed in
1935~\cite{Einstein:1935rr}. In brief, quantum entanglement implies
that a measurement performed in a physical system can give access to
information on another system at a different location {\it instantaneously}.
This quantum phenomenon is so intriguing that it is sometimes dubbed ``spooky
action at a distance". It was actually Erwin Schr$\rm\ddot{o}$dinger who
first clearly identified and named quantum entanglement in his papers
responding to the EPR paper~\cite{Schrodinger:1935zz,Schrodinger:1935oxy,
Schrodinger:1935flb,Schrodinger:1935ee} (see, e.g., Ref.~\cite{Price:2024zyu}
for a recent review). But quantum entanglement turned out to be true, and
the EPR paper turned out to be the most cited paper of Albert Einstein's
as recorded in the INSPIRE database. In comparison, Schr$\rm\ddot{o}$dinger's
cat as a vivid illustration of quantum entanglement became more
popular around the world.

Let us imagine that there are two identical boxes containing
two Schr$\rm\ddot{o}$dinger's cats that originate from the same dynamical
process but are simultaneously {\it dead} and {\it alive}. So the two
invisible cats are quantum-entangled or quantum-correlated, implying that
they cannot be simply represented as a product of the states
$|{\rm dead}\rangle$ and $|{\rm alive}\rangle$.
Instead, the two entangled cats should be described by the state
\begin{eqnarray}
\Psi^{}_{\rm cats} \equiv
\frac{1}{\sqrt 2} \Big[|{\rm dead}\rangle^{}_1 |{\rm alive}\rangle^{}_2
+ |{\rm alive}\rangle^{}_1 |{\rm dead}\rangle^{}_2\Big] \; ,
\label{1}
%     (1)
\end{eqnarray}
where box ``1" and box ``2" are indistinguishable from outside. Now the two
boxes are separated from each other at a distance, such that they
lose their original locality and hence are expected to be no longer
correlated. However, this expectation is not true. The key point
of quantum entanglement is that as soon as one box is opened to make sure of
the situation of the inside cat (dead or alive), the situation of the cat
inside the other box can be simultaneously fixed (alive or dead).

In particle physics, the pseudoscalar mesons $K^0$ and $\bar{K}^0$ were
the first pair of Schr$\rm\ddot{o}$dinger's cats, thanks to the pioneering
insight into $K^0$-$\bar{K}^0$ mixing via the one-loop (box-diagram)
weak interactions by Murray Gell-Mann and Abraham Pais~\cite{Gell-Mann:1955ipe}.
The $K^0$-$\bar{K}^0$ mixing effect was first observed in 1956, and later on
the similar quantum effects of $B^0_d$-$\bar{B}^0_d$ mixing (1987),
$B^0_s$-$\bar{B}^0_s$ mixing (2006) and $D^0$-$\bar{D}^0$ mixing (2007)
were also measured~\cite{BaBar:2014omp}. These meson-antimeson correlation
systems not only provided a distinctive playground for testing the standard
model (SM) of particle physics, but also offered an ideal playground for
verifying the EPR effect~\cite{KLOE:2006iuj,Belle:2007ocp,CLEO:2008kim,
BESIII:2020khq,BESIII:2025pod,Cheng:2025zcf,Shi:2025ggs}.

The most striking success in this regard should be an unambiguous
measurement of the Kobayashi-Maskawa (KM) phase in 2001, with the
help of quantum-entangled $B^0_d$ and $\bar{B}^0_d$ mesons produced on the
$\Upsilon (4S)$ resonance in the BaBar and Belle
experiments~\cite{BaBar:2001pki,Belle:2001zzw}. In fact, this phase is
the only observed fundamental phase parameter of the SM which characterizes
the strength of weak CP violation~\cite{Kobayashi:1973fv}.
As well known in quantum mechanics and quantum field
theories, {\it behind non-observable phases may exist an underlying symmetry,
while an observable phase is expected to point to symmetry breaking}. The
nontrivial KM phase is just a good example for the breaking of CP
symmetry or matter-antimatter symmetry in the microworld.

\section{How is a pair of entangled meson and antimeson formed}

In particle physics a quark and an antiquark are defined
to have the positive and negative intrinsic parities, respectively.
The parity of a composite system in its ground state is the product of
the parities of its constituents. That is why a pseudoscalar meson
and its antimeson have the negative parity; e.g.,
${\cal P}|K^0\rangle = -|K^0\rangle$ and
${\cal P}|\bar{K}^0\rangle = -|\bar{K}^0\rangle$, where ${\cal P}$
denotes the parity transformation operator. In comparison, the photon
$\gamma$ is a vector boson and hence has the negative intrinsic parity.

By definition, a charge conjugation transformation $\cal C$ converts each
particle into its antiparticle. So it changes the signs of all the
{\it internal} quantum numbers such as the electric charge, baryon
number and lepton number. As for the neutral mesons $K^0$ and $\bar{K}^0$,
we have ${\cal C} |K^0\rangle = |\bar{K}^0\rangle$ and
${\cal C} |\bar{K}^0\rangle = |K^0\rangle$. Therefore,
${\cal CP}|K^0\rangle = -|\bar{K}^0\rangle$ and
${\cal CP}|\bar{K}^0\rangle = -|K^0\rangle$ hold under the CP
transformations. Using $P^0$ to represent a pseudoscalar meson $K^0$,
$D^0$, $B^0_d$ or $B^0_s$ in general, one may similarly arrive at the
CP transformations ${\cal CP}|P^0\rangle = -|\bar{P}^0\rangle$ and
${\cal CP}|\bar{P}^0\rangle = -|P^0\rangle$.

Given an electron-positron collider running at the resonance of a vector
quarkonium state $X(q\bar{q})$ with $J^{\rm PC} = 1^{--}$
[i.e., $e^- + e^+ \to \gamma^* \to X(q\bar{q}$)]
%%%%%%%%%%%%%%%%%%%%%%%%%%%%%%%%%%%%%%%%%%%%%%%%%%%%%%%%%%%%%%%%%%
\footnote{The reaction $e^- + e^+ \to Z^* \to X(q\bar{q})$ may also
take place, but its rate is expected to be highly suppressed for
$m^{}_X \ll m^{}_Z$ and thus negligibly small.},
%%%%%%%%%%%%%%%%%%%%%%%%%%%%%%%%%%%%%%%%%%%%%%%%%%%%%%%%%%%%%%%%%%
a pair of neutral meson $P^0$ and antimeson $\bar{P}^0$ with a definite
charge-conjugation parity C can be coherently produced from the strong
decay of this quarkonium.
Namely, $X(q\bar{q}) \to (P^0\bar{P}^0)^{}_{\rm C = -1}$
with the resulting $P^0\bar{P}^0$ pair in a P-wave state;
or $X(q\bar{q}) \to (P^0\bar{P}^0)^{}_{\rm C}
+ Y^{}_{\rm C^\prime}$, where $Y$ denotes a light particle or a combination
of two light particles with a definite $\rm C^\prime$, and
$\rm C^\prime = -C$ must hold to coincide with the initial quantum number
$\rm C = -1$ of $X(q\bar{q})$. In such a case,
the $(P^0\bar{P}^0)^{}_{\rm C}$ pair forms a quantum-entangled eigenstate
of both charge conjugation and parity,
\begin{eqnarray}
\Psi^{}_{\rm C} \equiv \frac{1}{\sqrt 2} \Big[|P^0({\bf p})\rangle
|\bar{P}^0({\bf -p})\rangle + {\rm C} |\bar{P}^0({\bf p})\rangle
|P^0({\bf -p})\rangle\Big] \; ,
\label{2}
%     (2)
\end{eqnarray}
which satisfies ${\cal C} \Psi^{}_{\rm C} = {\rm C} \Psi^{}_{\rm C}$
and ${\cal P} \Psi^{}_{\rm C} = {\rm C} \Psi^{}_{\rm C}$
with ${\rm C} = \pm 1$. As a consequence, we are left with
${\cal CP} \Psi^{}_{\rm C} = \Psi^{}_{\rm C}$. In view of
$P^0$-$\bar{P}^0$ mixing, the mass eigenstates $P^{}_{\rm H}$ (heavy)
and $P^{}_{\rm L}$ (light) of $P^0$ and $\bar{P}^0$ mesons
can be written as~\cite{Branco:1999fs,Bigi:2000yz}
\begin{eqnarray}
|P^{}_{\rm H}\rangle & \equiv & p |P^0\rangle + q |\bar{P}^0\rangle \; ,
\nonumber \\
|P^{}_{\rm L}\rangle & \equiv & p |P^0\rangle - q |\bar{P}^0\rangle \; ,
\label{3}
%     (3)
\end{eqnarray}
where $|p|^2 + |q|^2 = 1$ holds
%%%%%%%%%%%%%%%%%%%%%%%%%%%%%%%%%%%%%%%%%%%%%%%%%%%%%%%%%%%%%%%%%%%%
\footnote{In the limit of CP conservation with $p = q = 1/\sqrt{2}$,
$|P^{}_{\rm H}\rangle$ and $|P^{}_{\rm L}\rangle$ turn out to be
the odd and even CP eigenstates $|P^{}_-\rangle \equiv
\left(|P^0\rangle + |\bar{P}^0\rangle\right)/\sqrt{2}$ and
$|P^{}_+\rangle \equiv \left(|P^0\rangle - |\bar{P}^0\rangle\right)
/\sqrt{2}$, respectively.}.
%%%%%%%%%%%%%%%%%%%%%%%%%%%%%%%%%%%%%%%%%%%%%%%%%%%%%%%%%%%%%%%%%%%%
It is therefore straightforward to arrive at
\begin{eqnarray}
\Psi^{}_{\rm C} & \propto & \Big\{
\left(1 + {\rm C}\right) \Big[|P^{}_{\rm H}({\bf p})\rangle
|P^{}_{\rm H}({\bf -p})\rangle - |P^{}_{\rm L}({\bf p})\rangle
|P^{}_{\rm L}({\bf -p})\rangle\Big]
\nonumber \\
& & \hspace{-0.03cm} - \left(1 - {\rm C}\right)
\Big[|P^{}_{\rm H}({\bf p})\rangle
|P^{}_{\rm L}({\bf -p})\rangle - |P^{}_{\rm L}({\bf p})\rangle
|P^{}_{\rm H}({\bf -p})\rangle\Big] \Big\} \; .
\label{4}
%     (4)
\end{eqnarray}
We observe that the components $|P^{}_{\rm H}({\bf p})\rangle
|P^{}_{\rm H}({\bf -p})\rangle$ and $|P^{}_{\rm L}({\bf p})\rangle
|P^{}_{\rm L}({\bf -p})\rangle$ disappear in the ${\rm C} = -1$ case,
simply because the possibility of two identical bosons in an antisymmetric
P-wave state is absolutely forbidden by Bose-Einstein statistics.

The explicit forms of $X(q\bar{q})$ include $\phi(s\bar{s})$ to produce
coherent $K^0\bar{K}^0$ at a $\phi$ factory, $\psi(c\bar{c})$ to produce
coherent $D^0\bar{D}^0$ at a $\tau$-charm factory, and $\Upsilon(b\bar{b})$
which can be either $\Upsilon({\rm 4S})$ to produce coherent $B^0_d\bar{B}^0_d$
or $\Upsilon({\rm 5S})$ to produce coherent $B^0_s\bar{B}^0_s$ at a $B$
factory. Taking the charmonium threshold for example, one
has~\cite{BESIII:2025pod,Xing:1996pn,Shi:2016bvo,Friday:2025gpj}
\begin{eqnarray}
e^+ e^- \to \gamma^* \to \left\{
\begin{array}{l}
(D^0 \bar{D}^0)^{}_{\rm C = -1} \; , \\ \vspace{-0.3cm} \\
D^{0 *} \bar{D}^0 + D^0 \bar{D}^{0 *} \to \gamma +
(D^0 \bar{D}^0)^{}_{\rm C = +1} \; , \\ \vspace{-0.3cm} \\
D^{0 *} \bar{D}^0 + D^0 \bar{D}^{0 *} \to \pi^0 +
(D^0 \bar{D}^0)^{}_{\rm C = -1} \; , \\ \vspace{-0.3cm} \\
D^{0 *} \bar{D}^{0 *} \to \gamma\pi^0 +
(D^0 \bar{D}^0)^{}_{\rm C = +1} \; , \\ \vspace{-0.3cm} \\
D^{0 *} \bar{D}^{0 *} \to \gamma\gamma +
(D^0 \bar{D}^0)^{}_{\rm C = -1} \; , \\ \vspace{-0.3cm} \\
D^{0 *} \bar{D}^{0 *} \to \pi^0\pi^0 +
(D^0 \bar{D}^0)^{}_{\rm C = -1} \; .
\end{array}
\right.
\label{5}
%     (5)
\end{eqnarray}
A very recent exciting news is that the BESIII Collaboration has for
the first time observed the quantum correlation of $D^0$ and
$\bar{D}^0$ mesons with $\rm C = +1$~\cite{BESIII:2025pod}.

\begin{figure}[t]
\centerline{\includegraphics[width=10.5cm]{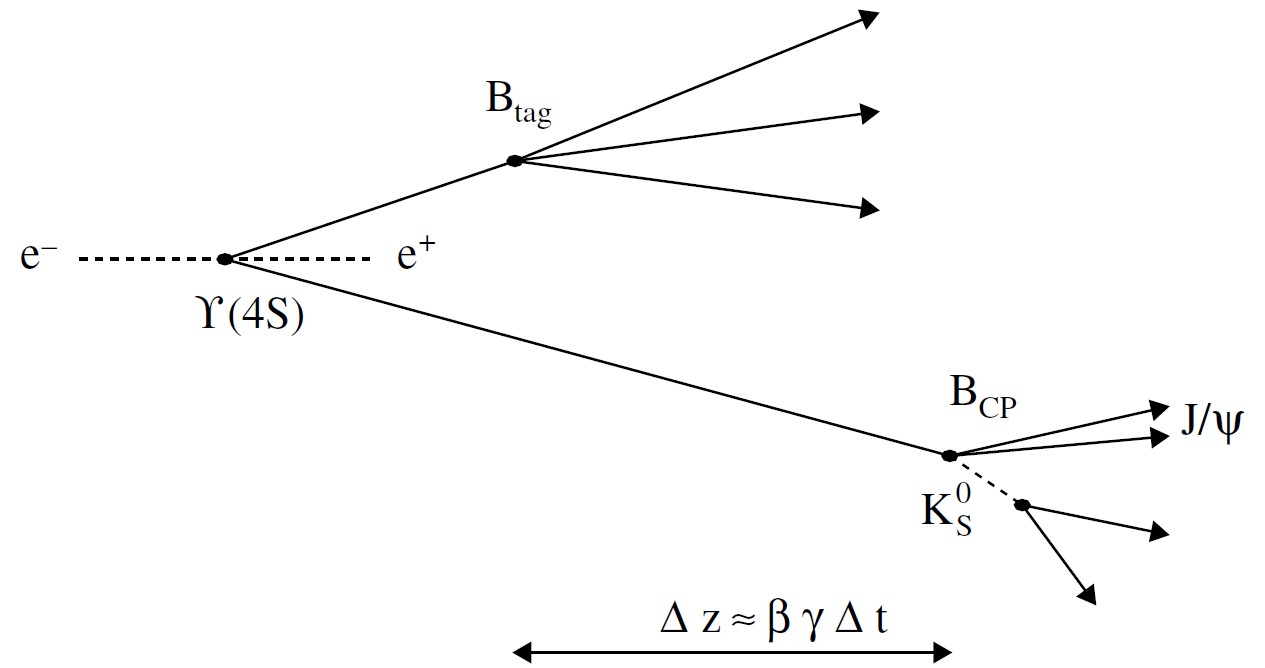}}
\caption{A schematic illustration of the geometry for an asymmetric
electron-positron collision to produce the
$\Upsilon({\rm 4S}) \to B^0_d \bar{B}^0_d$ events, where the decay of
one neutral $B$ meson serves for flavor tagging and the decay of the
other $B$ meson is to observe CP violation~\cite{BaBar:2014omp}.}
\label{Figure}
\end{figure}

While most of the electron-positron colliders are symmetric in
the sense that the electron and positron beams have the same energy
but the opposite momenta, the PEP-II and KEKB $B$-meson factories
are asymmetric to separate the vertices of the decaying $B^0_d$ and
$\bar{B}^0_d$ mesons which have been coherently produced on the
$\Upsilon ({\rm 4S})$ resonance. The main purpose of such an asymmetric
$e^+ e^-$ collision is to perform a time-dependent
measurement of CP-violating asymmetries in neutral $B$-meson decays.
The reason is simply that $B^0_d$ and $\bar{B}^0_d$ mesons must have
decay lengths in the laboratory which are long enough to assure that
the time sequence of their decays can be measured to a good degree of
accuracy, as schematically illustrated in Fig.~\ref{Figure}, where
the decay of one neutral $B$ meson serves for flavor tagging and the
decay of the other quantum-entangled $B$ meson is to probe
weak CP violation~\cite{BaBar:2014omp}.

It was Piermaria Oddone who proposed the original idea of making use
of an asymmetric $B$-meson factory to study CP violation in 1987.
This novel idea not only led to the successful constructions of PEP-II
at the SLAC (with the beam energies $E^{}_- = 9.0 ~{\rm GeV}$ and
$E^{}_+ = 3.1 ~{\rm GeV}$, together with the Lorentz boost factor
$\beta\gamma = 0.56$) and KEKB at the KEK ($E^{}_- = 8.0 ~{\rm GeV}$,
$E^{}_+ = 3.5 ~{\rm GeV}$ and $\beta\gamma = 0.425$) in
1999~\cite{BaBar:2014omp}, but also led Oddone to the 2005 APS Panofsky
Prize.
%%%%%%%%%%%%%%%%%%%%%%%%%%%%%%%%%%%%%%%%%%%%%%%%%%%%%%%%%%%%%%%%%%%%%
%%%%%%%%%%%%%%%%%%%%%%%%%%%%%%%%%%%%%%%%%%%%%%%%%%%%%%%%%%%%%%%%%%%%%

Let us mention that the mass eigenstates $B^{}_{\rm H}$ and
$B^{}_{\rm L}$ can be coherently produced on the $\Upsilon (4{\rm S})$
resonance at a $B$-meson factory, as shown in Eq.~(\ref{4})
with $\rm C = -1$. If CP were conserved in the $B^0_d$-$\bar{B}^0_d$
system, it would be reasonable to assume $B^{}_{\rm H} = B^{}_-$ and
$B^{}_{\rm L} = B^{}_+$ with $B^{}_{\pm}$ being the CP-even and CP-odd
states of $B^0_d$ and $\bar{B}^0_d$ mesons. In such a hypothetical
case the joint decay mode $B^{}_+ B^{}_- \to f^{}_{\rm CP} f^{}_{\rm CP}$
with $f^{}_{\rm CP}$ being a CP eigenstate would be
forbidden~\cite{Wolfenstein:1983cx,Gavela:1985dw,Bigi:1986vr}, since
$B^{}_+ \to f^{}_{\rm CP}$ and $B^{}_- \to f^{}_{\rm CP}$ would have no
way to simultaneously take place. In practice, however, the reaction
$B^{}_{\rm H} B^{}_{\rm L} \to f^{}_{\rm CP} f^{}_{\rm CP}$ is possible
to occur as a consequence of CP violation~\cite{Xing:1994mn}, although
the corresponding decay rate is expected to be highly suppressed.

\section{A decisive test of the KM mechanism of CP violation}

After the first discovery of CP violation in neutral $K$-meson decays
by James Cronin and Val Fitch's team in 1964~\cite{Christenson:1964fg},
a number of mechanisms of CP violation were proposed. Among them, the
most convincing one was the KM mechanism as it was based on the
renormalizable electroweak theory (i.e., the Weinberg
model~\cite{Weinberg:1967tq}) extended with three families of
quarks~\cite{Kobayashi:1973fv} and attributed the origin of flavor mixing
and CP violation to the co-existence of the dynamically distinctive
interactions of quark fields with the Higgs and gauge fields. That is why the KM mechanism
should not be regarded as a trivial extension of the Cabibbo flavor mixing
scenario~\cite{Cabibbo:1963yz}.

In the mass basis of six quark fields, it is the $3\times 3$ unitary
Cabibbo-Kobayashi-Maskawa (CKM) matrix $V$ that describes flavor mixing
and CP violation in the weak charged-current interactions of the SM:
\begin{eqnarray}
-{\cal L}^{}_{\rm cc} = \frac{g}{\sqrt{2}} \hspace{0.1cm} \overline{
\left(u ~~ c ~~ t \right)^{}_{\rm L}} \hspace{0.1cm} \gamma^\mu
\left(\begin{matrix}
V^{}_{ud} & V^{}_{us} & V^{}_{ub} \cr
V^{}_{cd} & V^{}_{cs} & V^{}_{cb} \cr
V^{}_{td} & V^{}_{ts} & V^{}_{tb} \cr\end{matrix}\right) \hspace{-0.1cm}
\left(\begin{matrix}
d \cr s \cr b \cr\end{matrix}\right)_{\rm L} W^+_\mu
+ {\rm h.c.} \; ,
\label{6}
%     (6)
\end{eqnarray}
where the nine elements of $V$ involve an {\it irremovable} phase which
violates the CP symmetry of weak interactions in the SM. There are many
ways to express such a fundamental phase parameter of
$V$~\cite{Fritzsch:1999ee,Xing:2020ijf}, but here we only discuss two
of them.
\begin{itemize}
\item     A conventional approach: introducing a physical phase parameter
of the CKM matrix $V$ in terms of a ``quartet" of its nine elements involving
four different quark flavor indices, such as
\begin{eqnarray}
\beta \equiv \arg\left(-\frac{V^{}_{cd} V^*_{cb}}{V^{}_{td}
V^*_{tb}}\right) \; .
\label{7}
%     (7)
\end{eqnarray}
This phase is actually one of the inner angles of the CKM unitarity
triangle defined by $V^{}_{ud} V^*_{ub} + V^{}_{cd} V^*_{cb} +
V^{}_{td} V^*_{tb} = 0$ in the complex plane.

\item     A new approach: introducing a physical phase parameter of
$V$ in terms of a ``trio" of its elements involving all the
six quark flavor indices, such as
\begin{eqnarray}
\varphi^{ijk}_{\alpha\beta\gamma} \equiv \arg\left(\frac{V^{}_{\alpha i}
V^{}_{\beta j} V^{}_{\gamma k}}{\det V}\right) \; ,
\label{8}
%     (8)
\end{eqnarray}
where the Greek and Latin subscripts run respectively over
$(u, c, t)$ and $(d, s, b)$ with $\alpha \neq \beta \neq \gamma$
and $i \neq j \neq k$.
\end{itemize}
Note that both $\beta$ and $\varphi^{ijk}_{\alpha\beta\gamma}$ are
rephasing-invariant in the sense that they are
independent of any redefinitions of the phases of six quark fields, and
thus they are the experimentally observable phase parameters. As $\beta \to 0$
or $\pi$ would imply the collapse of all the CKM unitarity triangles,
it {\it does} serve as a characteristic measure of weak CP violation in the
quark sector. So does the phase $\varphi^{dsb}_{uct}$, for example.

It was Ashton Carter and Anthony Sanda who first discussed the possibilities
of exploring CP violation in $B$-meson decays~\cite{Carter:1980hr,Carter:1980tk},
but it was Ikaros Bigi and Sanda who discovered the ``golden channels" to test
the KM mechanism of CP violation: $B^0_d$ versus $\bar{B}^0_d \to
J/\psi K^{}_{\rm S}$~\cite{Bigi:1981qs}. These two decay modes involve
$B^0_d$-$\bar{B}^0_d$ mixing, direct $b$-quark
decays and $K^0$-$\bar{K}^0$ mixing, and hence the interplay of these three
effects leads to a clean observable CP-violating asymmetry characterized
by~\cite{Du:1986ia,Xing:2000cg}
\begin{eqnarray}
{\cal A}^{}_{\rm CP} = -{\rm Im}\left(\frac{q^{}_B}{p^{}_B} \cdot
\frac{V^{}_{cb} V^*_{cs}} {V^*_{cb} V^{}_{cs}} \cdot
\frac{q^{*}_K}{p^{*}_K}\right)
\simeq \sin 2\beta \; ,
\label{9}
%     (9)
\end{eqnarray}
where the minus sign is associated with ${\cal CP}|J/\psi K^{}_{\rm S}\rangle
= - |J/\psi K^{}_{\rm S}\rangle$, and the ratios
$q^{}_B/p^{}_B \simeq \left(V^{}_{td} V^*_{tb}\right)/\left(V^*_{td}
V^{}_{tb}\right)$ and $q^{}_K/p^{}_K \simeq \left(V^{}_{cd} V^*_{cs}\right)
/\left(V^*_{cd} V^{}_{cs}\right)$ hold as two excellent approximations for the
box-diagram contributions to the $B^0_d$-$\bar{B}^0_d$ and $K^0$-$\bar{K}^0$
mixing effects which are dominated respectively by the intermediate top and
charm quarks, and those uncertainties originating from the hadronic matrix
elements of $B^0_d$ and $\bar{B}^0_d$ decays into $J/\psi K^{}_{\rm S}$
have essentially been cancelled out.

A reliable measurement of $\sin 2\beta$, with the help of quantum entanglement
of neutral $B$ mesons on the $\Upsilon (4 {\rm S})$ resonance,
was reported by the BaBar and Belle
Collaborations almost at the same time in 2001~\cite{BaBar:2001pki,Belle:2001zzw}
%%%%%%%%%%%%%%%%%%%%%%%%%%%%%%%%%%%%%%%%%%%%%%%%%%%%%%%%%%%%%%%%%%%%%%%%%%%%%%%%%
\footnote{Today the experimental result is
$\sin 2\beta = 0.709 \pm 0.011$~\cite{ParticleDataGroup:2024cfk}.}.
%%%%%%%%%%%%%%%%%%%%%%%%%%%%%%%%%%%%%%%%%%%%%%%%%%%%%%%%%%%%%%%%%%%%%%%%%%%%%%%%%
As a happy consequence, Bigi and Sanda received the 2004 APS Sakurai Prize
for their pioneering work on the golden channel of $B$-meson decays towards
a clean test of the KM mechanism of CP violation
%%%%%%%%%%%%%%%%%%%%%%%%%%%%%%%%%%%%%%%%%%%%%%%%%%%%%%%%%%%%%%%%%%%%%%%%%%%%%%%%
\footnote{Carter quitted particle physics soon after his first postdoctoral
career in collaboration with Sanda at the Rockefeller University. He became a
politician and served as the 25th US Secretary of Defence under President Barack
Obama from February 2015 to January 2017.},
%%%%%%%%%%%%%%%%%%%%%%%%%%%%%%%%%%%%%%%%%%%%%%%%%%%%%%%%%%%%%%%%%%%%%%%%%%%%%%%%
while Makoto Kobayashi and Toshihide Maskawa received the 2008 Nobel Prize in Physics
for their pioneering prediction for the unique CP-violating phase parameter in weak
interactions of the SM.

\section*{Acknowledgments}

I am greatly indebted to Prof. Yu Shi for providing me a valuable opportunity to
talk at the Shanghai Workshop on Quantum Entanglement of High Energy Physics held
on 18 to 23 July 2025. I am also grateful to Prof. Haibo Li and Dr. Xiaoyu Li
for very useful discussions. This research was supported in part by the Scientific
and Technological Innovation Program of the Institute of High Energy Physics
under Grant No. E55457U2.

%%%%%%%%%%%%%%%%%%%%%%%%%%%%%%%%%%%%%%%%%%%%%%%%%%%%%%%%%%%%%%%%%%%%%%%%%%%%%%%%%%%%

\end{document}